\documentstyle[12pt,epsfig]{article}

\def\I{{\cal I}}
\def\F{{\cal F}}
\def\M{{\cal M}}

\date{}
\begin{document}
\title{{\LARGE }
Zero Temperature Dynamics of $2D$ and $3D$ Ising Ferromagnets}
\author{
{\bf Palani Sundaramurthy}\thanks{Department of Electrical and
Computer
Engineering, University of Arizona, Tucson, AZ 85721, USA}\\
\and {\bf D. L. Stein}\thanks{Departments of Physics and
Mathematics, University
of Arizona, Tucson, AZ 85721, USA}\\
} \maketitle
\begin{abstract}
We consider zero-temperature, stochastic Ising models $\sigma^t$ with
nearest-neighbor interactions in two and three dimensions.  Using both
symmetric and asymmetric initial configurations $\sigma^0$, we study the
evolution of the system with time. We examine the issue of convergence of
$\sigma^t$ and discuss the nature of the final state of the system. By
determining a relation between the median number of spin flips per site
$\nu$, the probability $p$ that a spin in the initial spin configuration
takes the value $+1$, and lattice size $L$, we conclude that in two and
three dimensions, the system converges to a frozen (but not necessarily
uniform) state when $p\ne1/2$. Results for $p=1/2$ in three dimensions are
consistent with the conjecture that the system does not evolve towards a
fully frozen limiting state. Our simulations also uncover `striped' and
`blinker' states first discussed by Spirin \emph{et al}.~\cite{Redner2},
and their statistical properties are investigated.
\end{abstract}

\section{Introduction}
\label{sec:intro}

Consider the stochastic process $\sigma^t$ corresponding to the
zero-temperature limit of Glauber dynamics for a ferromagnetic
Ising model with Hamiltonian
\begin{equation}
\label{eq:EA} {\cal H}= -\sum_{\langle x,y\rangle}\sigma_x
\sigma_y\ .
\end{equation}
Here $\langle\cdot\rangle$ denotes a sum over nearest neighbor
sites only and $x,y\in{\bf Z}^d$, the $d$-dimensional hypercubic
lattice.  So $\sigma^t$ takes values in ${\cal S}=\{-1,+1\}^{{\bf
Z}^d}$, the space of infinite-volume spin configurations on ${\bf
Z}^d$.  The process $\sigma^t$ can be interpreted as modelling the
nonequilibrium dynamical evolution of a classical Ising
ferromagnet following a deep quench.

Perhaps the most basic question that can be asked is whether the
process $\sigma^t$ settles down to a limit as $t\to\infty$.
Knowledge of this fundamental long-time property is a prerequisite
for studying persistence properties~\cite{St,De,DHP,MH,MS,NS99a},
and useful for the understanding of domain formation and
evolution, spatial and temporal scaling properties, and related
questions (for a review, see Ref.~\cite{Bray}), as well as aging
phenomena~(see, e.g.,~\cite{aging1,aging2,aging3,aging4,FINS01}).

Given the fixed lattice type and dynamics under consideration, the
convergence of $\sigma^t$ can depend only on dimension $d$ and on
the starting spin configuration $\sigma^0$.  (Other lattice types
were considered in~\cite{NNS00,NS00}.)  Rigorous results exist for
symmetric $\sigma^0$ (i.e., for all $x$, $\sigma^0_x=\pm 1$ with
equal probability, independently of all other spins, corresponding
to a quench from infinite to zero temperature in zero external
field) in $d=1,2$; here every spin flips infinitely
often~\cite{NNS00,NS00}.  Numerical studies by Stauffer~\cite{St}
indicate that this result may hold up to $d=4$, beyond which
$\sigma^t$ could have a limit. There are few numerical
studies~\cite{Redner} of the long-time behavior of $\sigma^t$ 
{\it asymmetric\/} $\sigma^0$ (corresponding to a deep quench in
nonzero external field) in dimensions greater than one.

\begin{figure}
\label{fig:flow}
\centerline{\epsfig{file=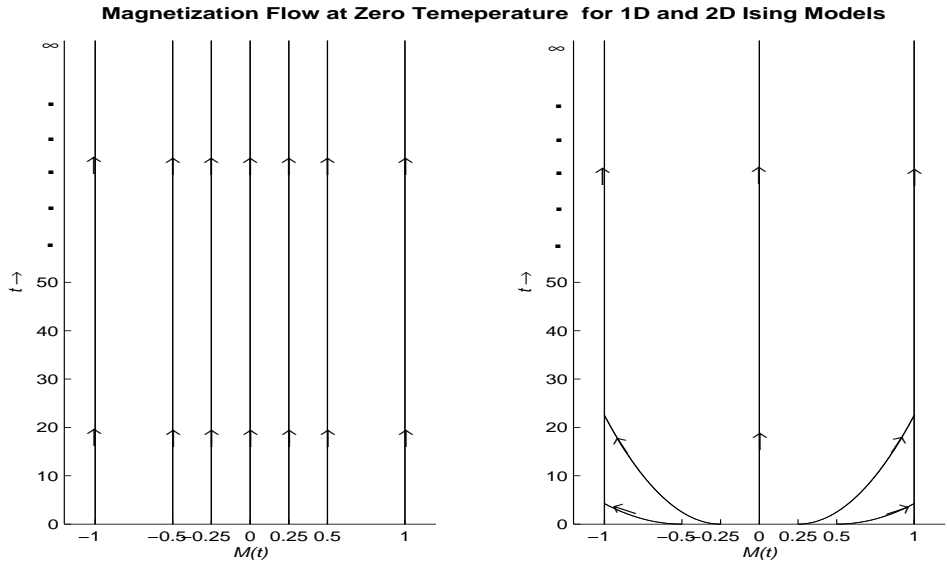,width=5.0in,height=3.0in}}
\renewcommand{\baselinestretch}{1.0}
\small \caption{Sketch of magnetization flow at zero temperature
under the Glauber dynamics discussed in the text.  (a)  One
dimension.  (b) Two dimensions.}
\end{figure}

In this paper we present the results of numerical studies that
examine the question of convergence of $\sigma^t$ for $d=2$ and
$3$ with both symmetric and asymmetric initial conditions. It is
known that the magnetization per spin at zero temperature
\begin{equation}
\label{eq:mag}
M(t)=\lim_{L\to\infty}{1\over|\Lambda_L|}\sum_{x\in\Lambda_L}\sigma_x(t)\,
,
\end{equation}
where $\Lambda_L$ is an $L^d$ cube centered at the origin, is
independent of time in one dimension for any initial
magnetization~\cite{Gl63}.  This time-invariance of the
magnetization under Glauber dynamics at zero temperature does not
hold in higher dimensions; indeed, it is easy to see, using the
independence of the spins in $\sigma^0$, that in $2D$ at time
$t=0$ the sign of $dM/dt$ is that of the initial magnetization.
For $d=2$ our results indicate that the long time behavior of the
symmetric case is unstable to small perturbations in the following
sense: for any deviation from up-down symmetry in the initial
state, $\sigma^t$ flows towards the uniform final state which is
magnetized in the direction of the initial asymmetry. While this
result is not surprising, it settles the question of whether the
nonexistence of $\sigma^\infty$ could persist for a nonzero range
of asymmetry in $\sigma^0$.  Our results for $2D$ are summarized
in Fig.~\ref{fig:flow}. For a $2D$ lattice of finite size, a
non-uniform final state called the frozen stripe state is also
possible. Such a configuration occurs when the spins are aligned
parallel to the axes and cease to flip thereafter.

\renewcommand{\baselinestretch}{1.25}

In addition to the frozen stripe state, a finite size $3D$ lattice
can evolve towards a `partially frozen' state (called `blinker
states' in~\cite{Redner}) when a subset of the spins continue to
flip ceaselessly, without any change in energy, while the
remainder of the spins do not flip any more. Our results for the
symmetric case in $3D$ agree with those of Stauffer~\cite{St}. For
$p=1/2$, we find, in agreement with~\cite{Redner}, frozen stripe
states and blinker configurations, to be discussed in the
following sections.

We also discuss the question of extrapolation of results on finite
size lattices to those of infinite extent.

\section{Definitions and Previous Results}
\label{sec:defs}

So far we have defined the problem informally: following a quench
from infinite to zero temperature, in zero or nonzero external
magnetic field, and subsequently evolving through standard Glauber
dynamics in zero field, will the spin configuration eventually
settle down to a final state, or will it continue to evolve
forever (and if so, in what sense)?

We now state the problem more precisely.  The initial spin
configuration $\sigma^0$ is chosen from a Bernoulli product
measure (denoted $P_{\sigma^0}$), with each spin independently
having value $+1$ with probability $p$ and $-1$ with probability
$1-p$.  The initial magnetization $M(0)=2p-1$.  The symmetric case
corresponds to $p=1/2$ and has $M(0)=0$; any other $P_{\sigma^0}$
will be called asymmetric.

The (single-spin-flip) dynamics will be taken to be continuous
time, given by independent rate 1 Poisson processes at each $x$
corresponding to those times $t_x$ (which can be thought of as
clock rings at $x$) when a spin flip
($\sigma_x^{t+0}=-\sigma_x^{t-0}$) is {\it considered\/}.  If the
resulting change in energy is negative, then the flip occurs with
probability 1, and if positive, with probability 0.  If there is a
``tie'', i.e., the resulting change in energy would be zero, then
the flip occurs with probability 1/2, as determined by a fair coin
toss.  We denote by $P_\omega$ the probability distribution on the
realizations $\omega$ of the dynamics.

So there are two sources of randomness, in the dynamics $\omega$
and in the initial spin configuration $\sigma^0$.  The joint
distribution of $\omega$ and $\sigma^0$ will be denoted $P$.  All
results to be discussed below are to be understood as occurring
with probability one in $P$.

The question posed in the introduction is whether $\sigma^t$ has a
limit, with $P$-probability one, as $t \to \infty$.  This is
equivalent to $\sigma_x^t$ flipping only finitely many times for
every $x$. Using the same nomenclature as in~\cite{NS00}, we call
such an $x$ an $\F$-site ($\F$ for finite); if $\sigma_x^t$ flips
infinitely often it is an $\I$-site ($\I$ for infinite).  By
translation-ergodicity, the collection of $\F$-sites (resp.,
$\I$-sites) has (with $P$-probability one) a well-defined
non-random spatial density $\rho_\F$ (resp., $\rho_\I$).  The
densities $\rho_\F$ and $\rho_\I$ $(=1-\rho_\F)$ can depend on
both $d$ and $p$.  A particular model will be said to be of type
$\F$, $\I$ or $\M$ (for mixed) according to whether $\rho_\F = 1$,
$\rho_\I = 1$ or $0<\rho_\F,\rho_\I<1$, respectively.

In one dimension the analysis is particularly simple, and it is
fairly straightforward to show that, for any $0<p<1$, the model is
type-$\I$~\cite{NNS00}.  (The proof given there is for $p=1/2$
only, but the result is not hard to extend to asymmetric models.)
In \cite{NNS00} it was also shown that the homogeneous ferromagnet
with $p=1/2$ on ${\bf Z}^2$ is type-$\I$; essentially the same
argument holds for the homogeneous antiferromagnet on ${\bf Z}^2$
with symmetric initial conditions.  No results were obtained for
asymmetric initial conditions.

We note for completeness that the $\pm J$ spin glass on ${\bf
Z}^2$ with symmetric initial conditions was shown to be type
$\M$~\cite{GNS00}; the proof is far more involved than those for
the homogeneous cases.  These are models in which couplings are
chosen {\it a priori\/} from a symmetric Bernoulli distribution,
with values $+1$ or $-1$, and are thereafter quenched.  (In fact a
much wider class of related models was also shown to be type-$\M$,
but these are less relevant to the models studied here.)

Under periodic boundary conditions, convergence to a final $+1$ or $-1$
state is not guaranteed: the lattice can form \emph{domain walls} parallel
to the lattice axes. These walls are stable under the dynamics (hence
referred to as the frozen `stripe' states), so any state with one or more
such walls (non-intersecting, i.e., all in either the $x$- or the
$y$-direction) is a final state.  A simple example of a domain wall in
a~$4\times4$ lattice is shown below:

\begin{center}
\begin{tabular}{l l l l}
$+$ & $+$ & $+$ & $+$\\
$+$ & $+$ & $+$ & $+$\\
$-$ & $-$ & $-$ & $-$\\
$-$ & $-$ & $-$ & $-$\\
\end{tabular}\\
\end{center}

Investigations into this freezing phenomenon for a zero-temperature Ising
ferromagnet having single-spin flip Glauber dynamics indicate that the
system need not always evolve into the uniform state.  Studies
investigating the final state of zero-temperature Ising
ferromagnets~\cite{Redner} by Spirin \emph{et al.}  indicate that when
$d=2$, the system reaches either a frozen `striped' state (i.e., with one
or more domain wall pairs) with probability $\approx$ $1/3$ or the uniform
state with probability $\approx$ $2/3$~\cite{HZS96}.  Moreover their study
indicates that these frozen states persist with positive probability as
$L\to\infty$. When $d>2$, persistent configurations other than the striped
or uniform state are also observed. In these, the system wanders through a
series of iso-energy partially frozen states. These metastable states are
called \emph{blinkers} in~\cite{Redner} and consist of localized sets of
spins that flip forever with zero energy cost, surrounded by frozen
spins. These blinkers become more numerous as the lattice size increases
and the system is more likely to be trapped in one of these
states. Presence of blinkers in the infinite system would indicate that it
is of type-$\M$. These results will be discussed further in the
Sec.~\ref{sec:conclusions}.

\section{Simulation Results}
\label{sec:sims}

Zero-temperature Glauber dynamics has been implemented by each
spin independently chosen to flip through a Poisson process having
rate parameter $\mu = 1$. When a particular site is selected, it
assumes the sign of the majority of its neighbors. In case of a tie,
the sign of the spin is determined by a fair coin toss.

\begin{figure}
\centerline{\epsfig{file=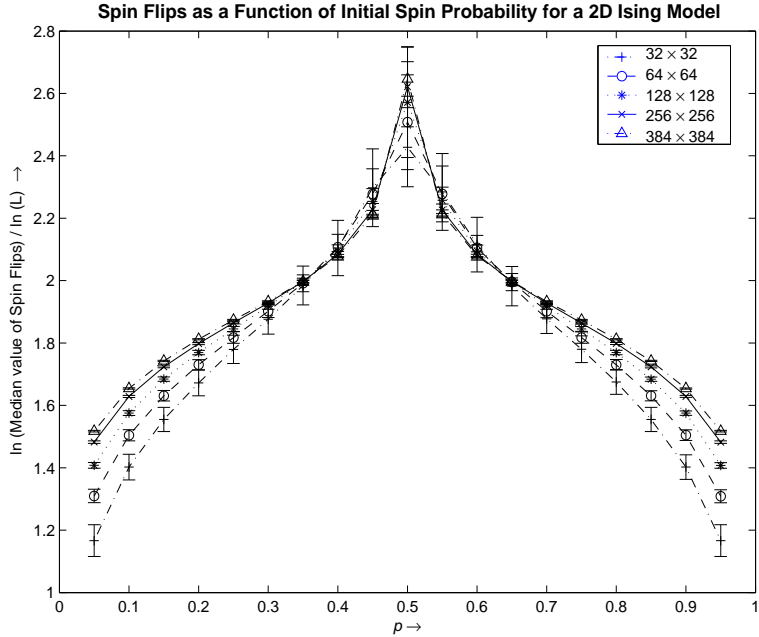,width=4.0in,height=3.4in}}
\renewcommand{\baselinestretch}{1.0}\small
\caption{$\ln{\cal N} /\ln L$, as a function of $p$ in a $2D$ Ising
Model. At least $10^3$ independent runs have been considered for $L
< 256$, and at least $5 \times 10^2$ independent runs for $L \ge
256$.} \label{fig:spinflips2d}
\end{figure}

\begin{figure}
\centerline{\epsfig{file=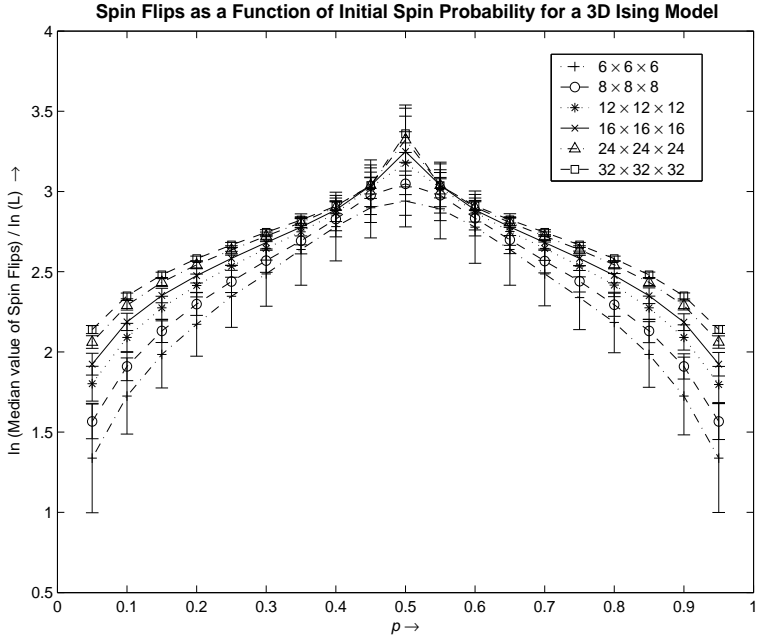,width=4.0in,height=3.4in}}
\renewcommand{\baselinestretch}{1.0} \small
\caption{$\ln{\cal N} /\ln L$, as a function of $p$ in a $3D$ Ising
Model. At least $10^3$ independent runs have been considered for all values
of $L$, and only configurations that settle down to a fully frozen final
state are included.} \label{fig:spinflips3d}
\end{figure}

Results depicting the median value of spin flips, ${\cal N}$, for $2D$ and
$3D$ Ising models of different lattice sizes are shown
in~Figs.~\ref{fig:spinflips2d} and~\ref{fig:spinflips3d} respectively. For
those runs in which the system converges to a final uniform state, ${\cal
N}$ has been computed by averaging several independent (both in choice of
initial spin configuration and in dynamical realization) runs.

Our simulations in $3D$ also uncovered blinker states. When such a state
appeared, the `unfrozen' spins (i.e. those that continued to flip) were
tracked.

From the simulation data for the $2D$ lattice, we observe that the number
of spin flips is symmetric about a maximum at $p = 1/2$ irrespective of the
lattice size. We also observe that the slope of this curve is sharpest near
$p = 1/2$, while it remains relatively constant for very small or large
values of $p$. For both the $2D$ and $3D$ models, the family of curves
corresponding to different lattice sizes collapses to a single curve when
$\ln{\cal N} /\ln{2}L$ is plotted vs.~$p$. For very large lattice sizes, we
have observed that the system almost always converges to a $\pm1$ state
even for small departures from $p = 1/2$.

\begin{figure}
\centerline{\epsfig{file=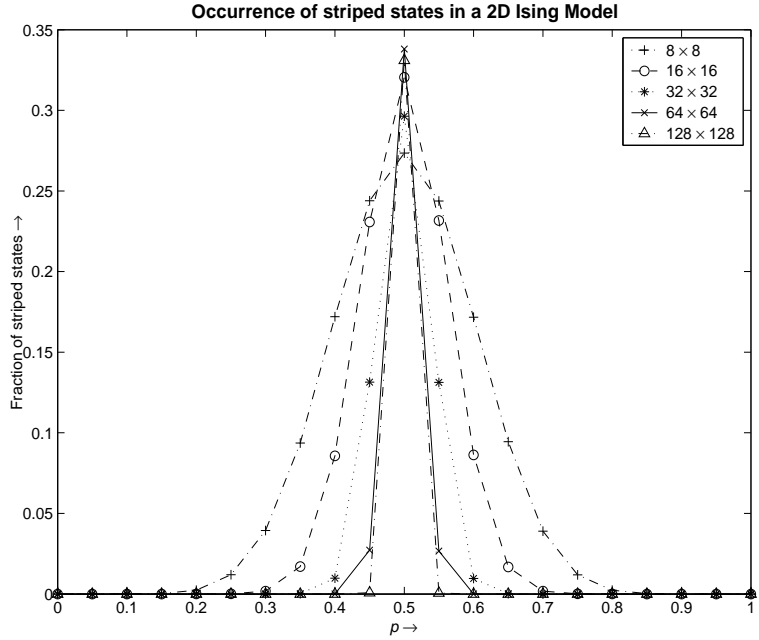,width=4.0in,height=3.4in}}
\renewcommand{\baselinestretch}{1.0} \small
\caption{Occurrence of striped states as a function of $p$ for a $2D$
Ising Model. At least $10^6$ independent runs have been considered for $p
\ne 1/2$, and at least $10^3$ runs for $p = 1/2$.}
\label{fig:domainwall2d}
\end{figure}

It is interesting to study the fraction of striped states as a function of
$p$ and lattice size $L$. This has been studied for a $2D$ model shown
in~Fig.~\ref{fig:domainwall2d}. We observe that when $p = 1/2$, as the
lattice size increases, the fraction of the spins converging to a final
striped state asymptotically tends towards $1/3$, in agreement with the
results in~\cite{Redner}. When $p \ne 1/2$, the fraction of final spin
configurations in the striped state decreases as lattice size increases;
i.e., the uniform state is increasingly favored. No blinker configurations
are present in the $2D$ lattice. These results are consistent with the
known result~\cite{NS00} that the infinite square lattice is type-$\I$ for
$p=1/2$, and they indicate that otherwise the lattice is type-$\F$.

\begin{figure}
\centerline{\epsfig{file=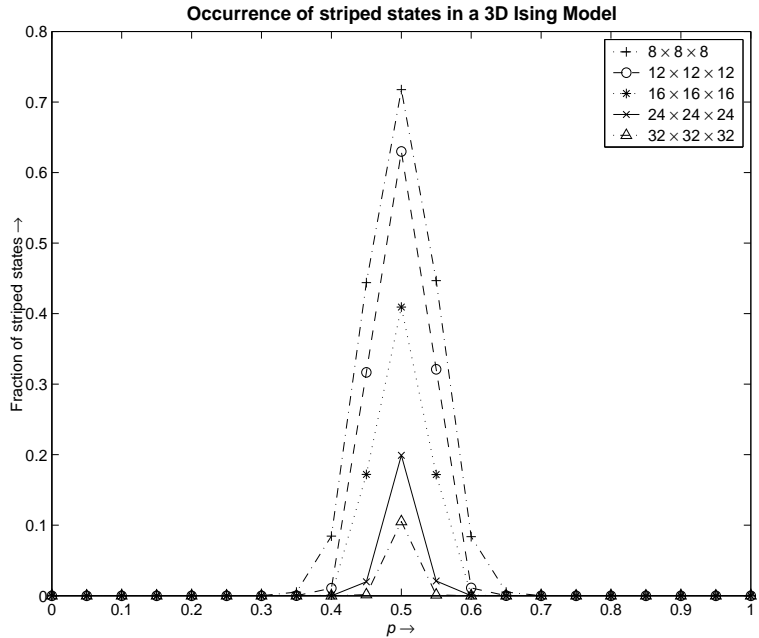,width=4.0in,height=3.4in}}
\renewcommand{\baselinestretch}{1.0}
\small \caption{Occurrence of striped states as a function of $p$ for
a $3D$ Ising Model. At least $10^6$ independent runs have been considered
for $p \ne 1/2$, and at least $10^4$ runs for $p = 1/2$.}
\label{fig:domainwall3d}
\end{figure}

\begin{figure}
\centerline{\epsfig{file=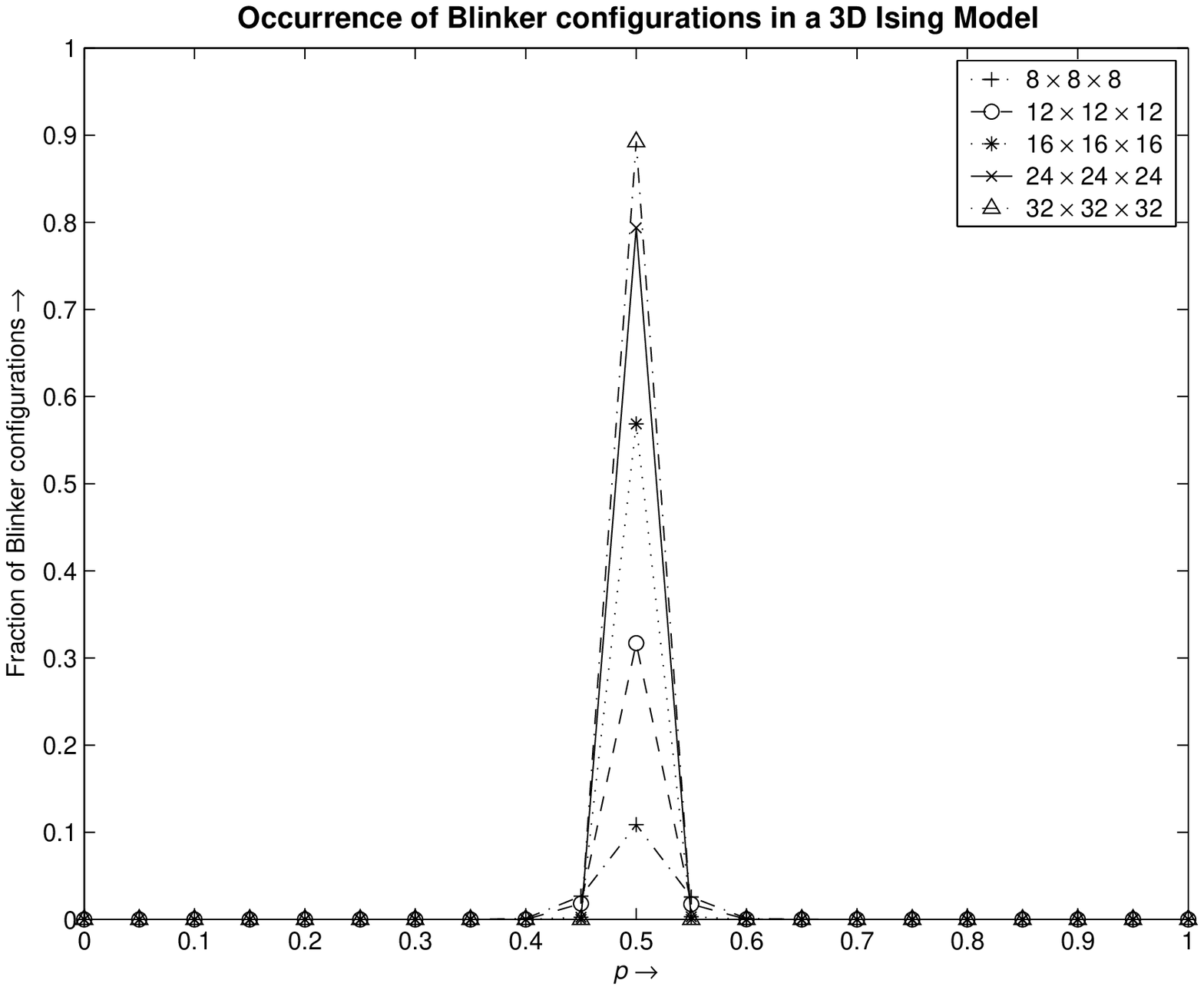,width=4.0in,height=3.4in}}
\renewcommand{\baselinestretch}{1.0}
\small \caption{Occurrence of blinker states as a function of $p$ for a
$3D$ Ising Model.  At least $10^6$ independent runs have been
considered for $p \ne 1/2$, and at least $10^4$ runs for $p = 1/2$.}
\label{fig:blinker3d}
\end{figure}

Fig.~\ref{fig:spinflips3d} shows results for the $3D$ cubic lattice; the
data collapse similarly to the $2D$ case. As expected, the maximum number
of spin flips occurs when $p = 1/2$.~Fig.~\ref{fig:domainwall3d} shows the
occurrence of striped states as a percentage of the total number of final
states while~Fig.~\ref{fig:blinker3d} depicts the occurrence of blinker
states. Comparison of Figs.~\ref{fig:domainwall3d} and \ref{fig:blinker3d}
clearly shows that, as $L$ increases, the fraction of blinker states
increases at the expense of the frozen striped states, in agreement
with~\cite{Redner2}.  These trends suggest that for larger lattices, even
small deviations from $p = 1/2$ causes the system to converge to the
uniform state, while for $p=1/2$, most final states are blinkers.  In the
latter case it seems likely that as $L\to\infty$, almost every final
configuration is a blinker state (but see the discussion in Section $4$).

\begin{figure}
\centerline{\epsfig{file=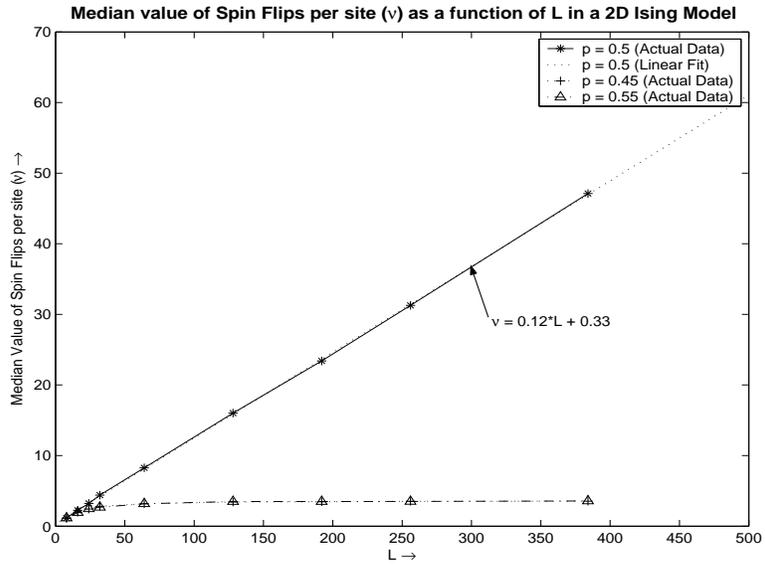,width=4.0in,height=3in}}
\renewcommand{\baselinestretch}{1.0}
\small \caption{Median value of spin flips per site, $\nu$, as a function of
$L$ in a $2D$ Ising Model.} \label{fig:nuL2d}
\end{figure}

\begin{figure}
\centerline{\epsfig{file=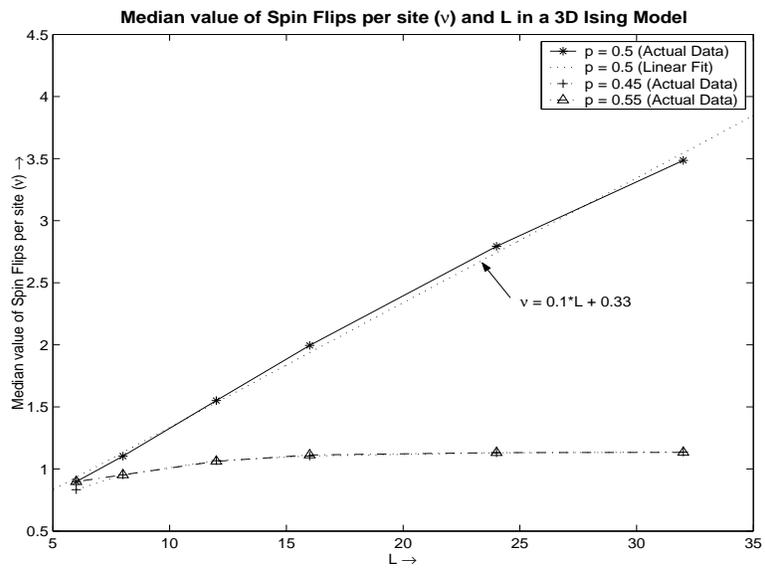,width=4.0in,height=3in}}
\renewcommand{\baselinestretch}{1.0}
\small \caption{Median value of spin flips per site, $\nu$, as a function of
$L$ in a $3D$ Ising Model.} \label{fig:nuL3d}
\end{figure}

The median number of spin flips per site, $\nu ={\cal N}/L^d$, was studied
in $2D$ and $3D$ models for different system sizes and different $p$'s
(0.45, 0.50, 0.55), when the system converged to either a striped or a
uniform state.  For $p = 0.45$ and $0.55$, $\nu$ increases with $L$ for
small system sizes, but rapidly levels off after (roughly) $L\sim100$ in
$2D$ and $L\sim20$ in $3D$.  Interestingly, these correspond to samples
with approximately equal numbers of spins.  Significantly, in both $2D$ and
$3D$ the rise of $\nu$ with $L$ continued steadily only at $p=1/2$, where
it increased monotonically as a linear function of $L$
(cf.~Figs.~\ref{fig:nuL2d} and~\ref{fig:nuL3d}) up to the largest sizes
considered. The rate of increase of $\nu$ with $L$ was, to within numerical
accuracy, virtually identical in $2D$ and $3D$.  This evidence indicates
that, in terms of the overall classification of long-time dynamical
behavior presented in Sec.~\ref{sec:defs}, the Ising ferromagnet on ${\bf
Z}^d$ behaves similarly for $d=2$ and $d=3$ as a function of $p$.  In
particular, both are type-${\cal F}$ when $p\ne 1/2$, while for $p=1/2$ the
$3D$ lattice is either type-${\cal I}$ as in $2D$~\cite{NNS00} or possibly
type-${\cal M}$.  These data, however, do not reveal the nature of the
final state(s) for large finite $L$; this will be discussed further below.

\begin{figure}
\centerline{\epsfig{file=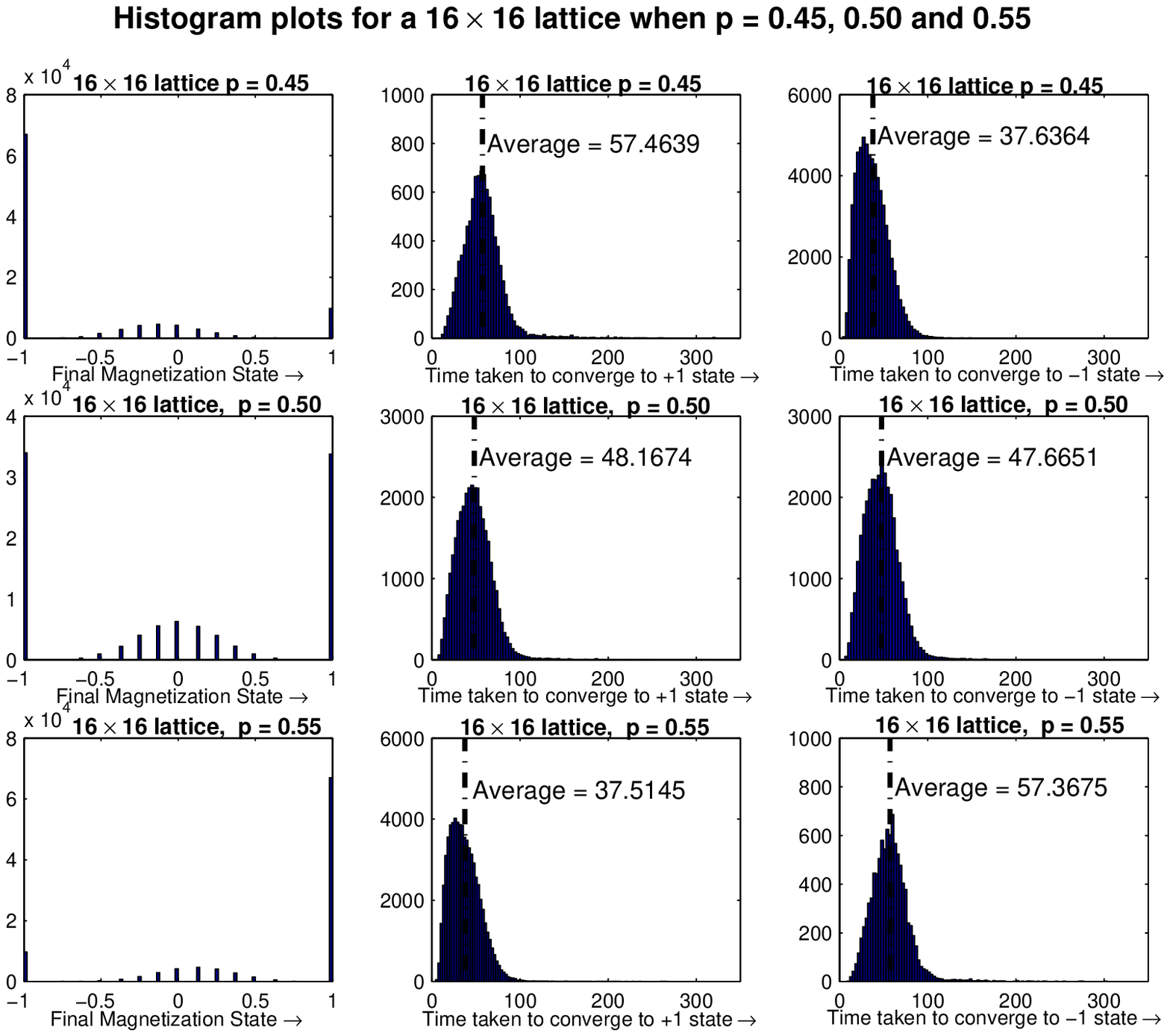,width=5.0in,height=5.5in}}
\renewcommand{\baselinestretch}{1.0}
\small \caption{Convergence time as computed by histogram plots for
the~$16\times 16$ lattice when $p = 0.45$, $0.50$ and $0.55$. In all
the plots the y-axis represents the frequency of occurrence. All the
histograms are divided into $100$ equally spaced bins and shown on a
linear scale. The first row shows final magnetization states and the
time taken to converge to the $-1$ and $+1$ states when $p = 0.45$.
The second and third rows show the results for $p = 0.50$ and $0.55$
respectively. The results along the first and third rows are similar
indicating symmetry around $p = 0.50$. } \label{fig:hist1616}
\end{figure}

We now study other aspects of the system evolution for $p$ near $1/2$. One
important quantity of interest is the convergence time to a final state
(for $p \ne 1/2$) as a function of $(p, d)$.  Fig.~\ref{fig:hist1616} shows
the result when $p=0.45$, $0.50$, and $0.55$, for a $16 \times16$ lattice.
Times shown are in units of $\mu^{-1}$, the inverse Poisson rate
parameter (here set equal to one).  Each simulation consists of $10^5$
independent runs where a final uniform or striped state was reached.

For the $2D$ Ising model, when $p=0.45$, $66.92\%$ of the runs have final
magnetization $-1$. In a smaller percentage of cases ($9.79\%$) the
minority spin prevails, i.e., the final magnetization is $+1$. The
remaining $23.29\%$ are in the striped state. The average time to reach the
$-1$ state is, not surprisingly, significantly smaller ($37.64$) than that
taken to reach the $+1$ state ($57.46$). This extra time corresponds to
large fluctuations that take the system from the majority to the minority
spin state. For the case when $p = 1/2$, the fraction of runs leading
respectively to the uniform $+1$ and $-1$ final states are remarkably close
($33.98\%$ and $33.76\%$, respectively) to the fraction $1/3$. The system
reaches a final striped state with a probability also close to $1/3$, in
good agreement with the results found by Spirin \emph{et
al}~\cite{Redner}. The corresponding time averages to reach these states
are likewise very close ($48.17$ and $47.67$, respectively). The results
for $p=0.55$ mimic those of $p=0.45$, with the relative roles of the $+1$
and $-1$ states switched. In this case $66.96\%$ reach the $+1$ final state
while $9.66\%$ reach the $-1$ state. The corresponding time averages are
$37.51$ and $57.37$, respectively.

\begin{figure}
\centerline{\epsfig{file=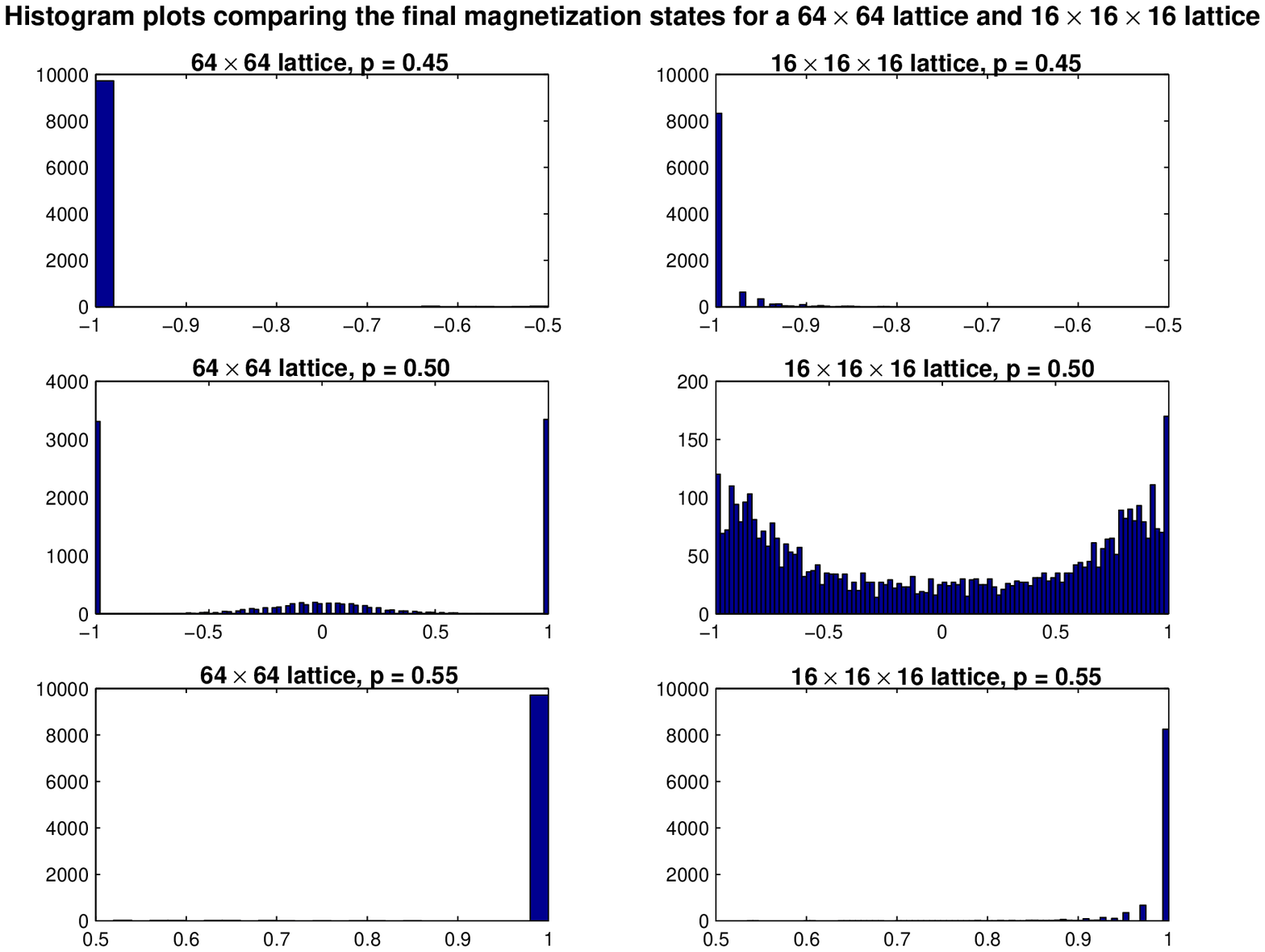,width=5.0in,height=5.5in}}
\renewcommand{\baselinestretch}{1.0}
\small \caption{Histogram plots comparing the final magnetization
states for a ~$64\times 64$ lattice with a ~$16\times 16\times 16$
lattice for $p = 0.45$, $0.50$ and $0.55$. In all plots the y-axis
represents the frequency of occurrence while the x-axis represents
the final magnetization states. All the histograms are divided
into $100$ equally spaced bins and shown on a linear scale. The
symmetry for the $p = 0.45$ and $0.50$ cases are evident from the
first and last rows of the graph.} \label{fig:histmagcomp}
\end{figure}

\begin{figure}
\centerline{\epsfig{file=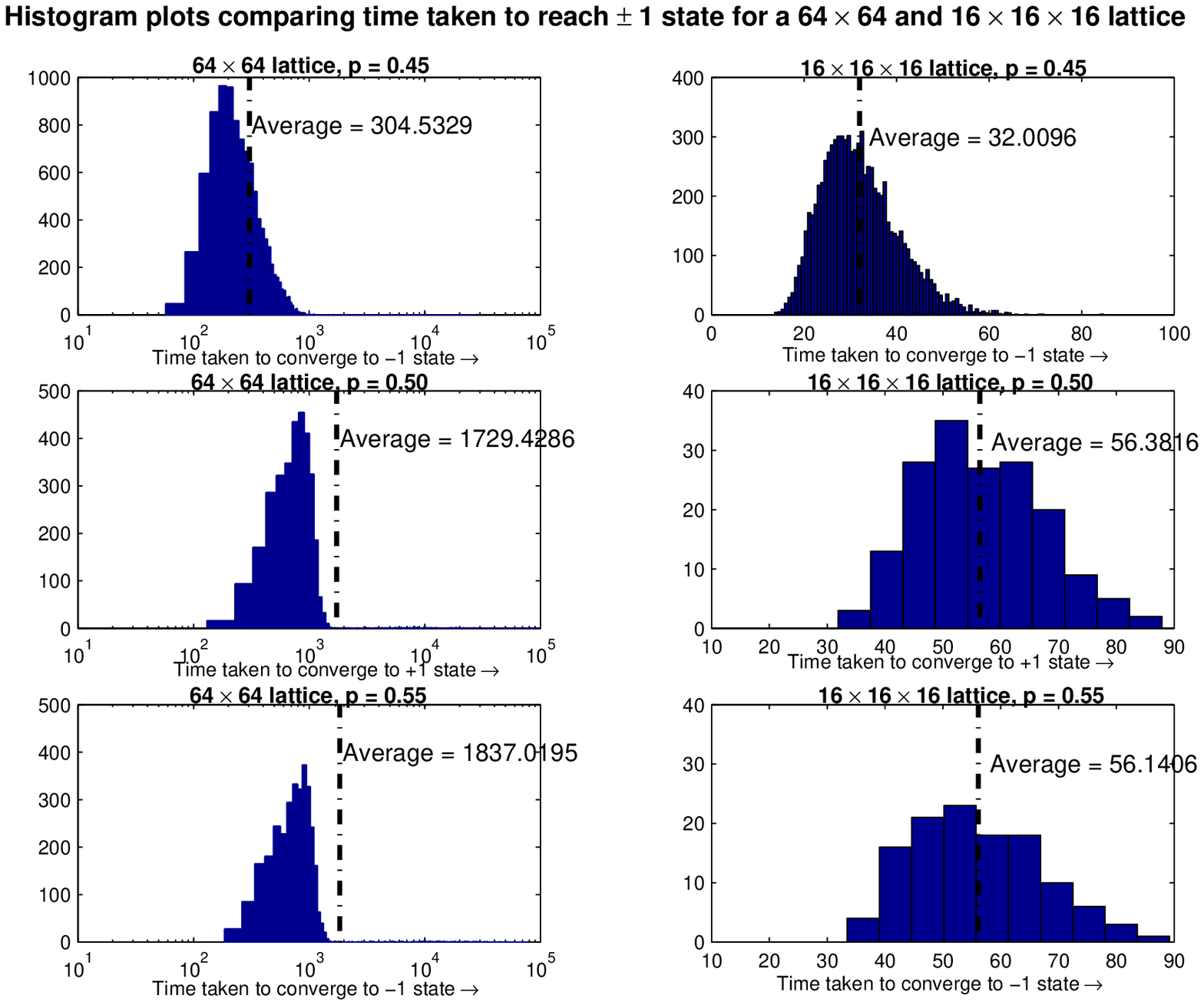,width=5.0in,height=5.5in}}
\renewcommand{\baselinestretch}{1.0}
\small \caption{Histogram plots comparing the time taken to reach
a final $\pm1$ state for a ~$64\times 64$ lattice with a
~$16\times 16\times 16$ lattice for $p = 0.45$, $0.50$ and $0.55$.
In all plots the y-axis represents the frequency of occurrence
while the x-axis represents the time taken. Note that the time
bins are depicted on a logarithmic scale for the $2D$ lattice and
linear scale for the $3D$ case. All the histograms are divided
into $100$ equally spaced bins.} \label{fig:histtimecomp}
\end{figure}

These tendencies are confirmed using larger lattices, as the next
few figures show. In these, we compare results between a ($2D$)
$64\times64$ lattice and a ($3D$) $16\times16\times16$ lattice,
which have the same total number of spins. The results are based on
$10^4$ independent runs for either case. Graphs displaying final
magnetization states and convergence times for the $2$ cases when $p
= 0.45$ and $p = 0.50$ are displayed in Figs.~\ref{fig:histmagcomp}
and~\ref{fig:histtimecomp} respectively.

Comparing these two cases, we observe that for the $3D$ Ising model,
whenever convergence occurs, there is an overwhelming tendency for the
majority spins to prevail in the final state. For $p=0.45$, none of the
$10^4$ runs in the $3D$ case converged towards the $+1$ state compared to
$11$ instances in the $2D$ case where the minority spin prevailed in the
final state. For a $64\times64$ lattice, $97.18\%$ converged to the $-1$
state, $0.11\%$ converged to the $+1$ state and the rest ($2.21\%$) were
striped states. Similarly, for the $16 \times 16 \times 16$ lattice,
$83.24\%$ converged to the $-1$ state, $16.54\%$ evolve into striped states
and the remainder ($0.11\%$) into blinker states. The average convergence
time is smaller and more predictable for the $3D$ model than for the $2D$
case. For the~$16\times 16\times 16$ lattice with $p=0.45$, the average
time taken to converge to the $-1$ state is $32.00$ compared to $304.53$
for the $64\times64$ lattice for the same $p$. When $p=0.50$, results are
symmetric, as expected. Average times taken to reach the $+1$ and $-1$
states in the ~$16\times16\times16$ lattice were $56.38$ and $56.14$,
respectively, while they were $1729.43$ and $1837.02$ for the~$64\times64$
lattice. Compared to the $2D$ case, there is a smaller fraction of the
spins converging to the final uniform state in the $3D$ case ($2.9\%$) and
this probability only decreases as $L$ increases. Also when $p=0.50$ in the
$3D$ lattice, configurations that have not converged to the uniform $+1$ or
$-1$ configurations tend to be closer to one of those two states than in
the $2D$ lattice (cf.~Fig.~\ref{fig:histmagcomp}).

\begin{figure}
\centerline{\epsfig{file=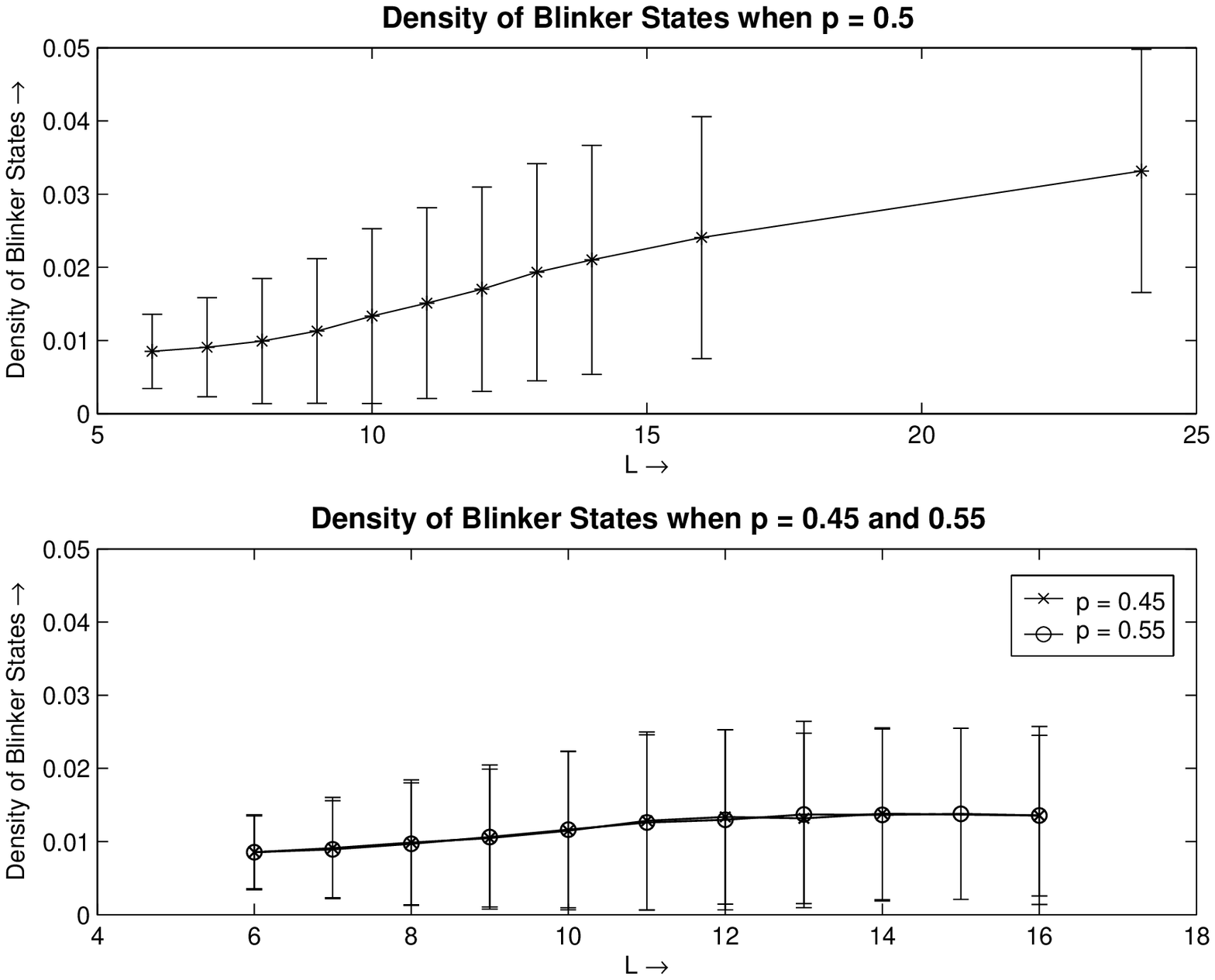,width=5.0in,height=4.5in}}
\renewcommand{\baselinestretch}{1.0} \small \caption{Plots showing the
blinker density (that is, the fraction of blinker spins in a given blinker
configuration) for a $3D$ lattice as a function of lattice size when $p =
0.45$, $0.50$ and $0.55$. In these plots the y-axis represents the the mean
number of blinker spins per blinker configuration, while the x-axis
represents lattice size.}
\label{fig:blinkerdensity}
\end{figure}

Fig.~\ref{fig:blinkerdensity} displays the nature of the blinker states in
$3D$.  Typically, the number and position of blinker spins fluctuates
greatly from one configuration to another, hence the relatively large error
bars. In these simulations, we considered at least $10^4$ runs for $L < 16$
and $10^3$ runs for $L \geq 16$. Two conclusions can be drawn from these
data: first, at the lattice sizes studied the density of blinker spins is
quite small; and second, there is a qualitative difference between the
symmetric and asymmetric cases.  In particular, when $p = 1/2$ the blinker
density {\it may\/} be increasing with lattice size $L$, while it appears
to remain flat for asymmetric initial conditions.  More work needs to be
done for larger $L$ to determine whether this trend persists.

\bigskip

\section{Discussion and Conclusions}
From our simulations for the evolution of the $2D$ and $3D$ Ising models
with random initial conditions and evolving under zero-temperature Glauber
dynamics, we draw the following conclusions:

\begin{itemize}

\item For $d = 2$, the median number of spin flips is symmetric about a
maximum at $p = 1/2$. The median number of spin flips per site, $\nu$,
increases with lattice size but the rate of increase is significant only
when $p = 1/2$~(cf.~Fig.~\ref{fig:nuL2d}), when $\nu = 0.12L + 0.33$. At
$p=1/2$ and large $L$, the system evolves respectively towards the uniform
$+1$ state, the uniform $-1$ state, and the striped state with equal
probability $1/3$ (over initial configurations and dynamical realizations).

\item However, it was proven in~\cite{NNS00} that the infinite lattice with
$p = 1/2$ is type-$\I$. This may suggest that the $L\to\infty$ limit, which
distinguishes between striped and uniform states in a clear limiting sense,
differs from the $L=\infty$ case. It is interesting to note that in the
infinite square, at almost every sufficiently large time, any large region
will {\it locally\/} resemble one of the two uniform phases or a domain
wall phase.  But this same region will continually approximate each of
these three possible states infinitely often, separated (probably) by
increasingly larger time intervals.  On the infinite lattice, this state of
affairs results from an unending supply of domain walls coming into any
fixed region `from infinity' --- a state of affairs that cannot occur for a
finite lattice, no matter how large.  These results and considerations
demonstrate that caution needs to be exercised in extrapolating numerical
results to infinite lattices in two or more dimensions.

\item For $d = 3$, we again found the median number of spin flips to be
symmetric about a maximum at $p = 1/2$. For this model the number of spin
flips per site again increased linearly with $L$: $\nu = 0.1L + 0.33$.  It
seems very likely that, as in the $2D$ case, a $3D$ lattice for $p = 1/2$
is not type-$\F$. The rate of increase of the median number of spin flips
per site is, within the accuracy of the simulations, close to that of the
$2D$ case.

As $L$ increases at $p=1/2$, the fraction of final spin configurations in
the blinker state increases rapidly (cf.~Fig.~\ref{fig:blinker3d}). For the
lattice sizes considered, however, the fraction of spins that are unfrozen
in each of these `blinker states' remains quite low. This is not surprising
--- it seems reasonable that every spin, or small connected cluster of
spins, that cycles endlessly through a small number of states requires a
significantly larger number of surrounding frozen spins to supply the
`boundary conditions' needed for this cycling to occur.

It is unclear at this stage whether, as $L$ increases, the fraction of
`unfrozen' spins in a typical final configuration levels off to a limiting
value strictly less than (and probably considerably smaller than) one, as
required for a blinker state, or continues to increase to one.  The former
case corresponds to type-${\cal M}$ and the latter to type-${\cal I}$.
Either way, these data are consistent with the conjecture that the
symmetric case in $3D$ is not type-${\cal F}$. Our data indicate that the
symmetric case {\it may\/} be type-$\M$ when $L=\infty$; however, the
previous example of the $2D$ case leaves open the possibility that the
$L\to\infty$ limit is type-${\cal M}$ while the $L=\infty$ case, i.e., the
infinite cubic lattice, is type-$\I$, and there is a corresponding
dynamical singularity in the $L\to\infty$ limit.

\item For $p\neq1/2$, our results strongly suggest that the system is
type-$\F$ when $L=\infty$ and $d=2$ or $3$, given that the average number
of spin flips per site does not appreciably change with increasing lattice
size.

\item Both $L$ and $d$ influence the probability of the minority
spin prevailing in the final converged state. For any given $p\ne 1/2$,
this probability decreases as $L$ or $d$ increases, thereby suggesting that
the occurrence of a final minority spin ground state has zero probability
when $L=\infty$.

\item As might be expected, the average time taken to converge to a final
majority state is smaller than that needed to converge to a minority
state. For the same number of sites and any given $p$, dynamics on the $3D$
lattice shows a tendency to converge faster than in the $2D$ case. When
$p=1/2$, the times taken to reach the $+1$ or $-1$ final state is almost the
same for both $2D$ and $3D$.

\end{itemize}

\label{sec:conclusions}

{\it Acknowledgments.\/} This work was partially supported by the
National Science Foundation under Grant DMS-01-02541.  The authors
thank C.~Newman and S.~Redner for useful discussions and
suggestions.

\renewcommand{\baselinestretch}{1.0}

\end{document}